\documentclass[a4paper,11pt]{article}

\usepackage{parskip}

\usepackage{graphicx}
\usepackage{amsmath}        
\usepackage{amssymb}        
\usepackage{amsfonts}       
\usepackage{amsbsy}     

\usepackage{hyperref}

\newcommand{\sina}{\sin\alpha}\newcommand{\cosa}{\cos\alpha}
\newcommand{\tb}{\tan\beta}
\newcommand{\ztwo}{\mathbb{Z}_2}
\newcommand{\hp}{{H^\pm}}
\newcommand{\ii}{\mathrm{i}}
\newcommand{\hcal}{ \mathcal{H} }
\newcommand*{\THDMC}{{\sc 2hdmc}}
\newcommand{\MadG}{{\sc MadGraph}}
\newcommand{\vev}[1]{ \left\langle {#1} \right\rangle }
\newcommand{\grp}[1]{\mathrm{#1}}
\newcommand{\FC}{{\sc FormCalc}}

\newcommand{\reffig}[1]{{Fig.~\ref{#1}}}

\newcommand{\eref}[1]{(\ref{#1})}

\newcommand{\be}{\begin{equation}}
\newcommand{\ee}{\end{equation}}

\begin{document}
\begin{LARGE}
\textbf{Higgs properties in the Stealth Doublet Model}\footnote{arXiv version of LHCP 2013 conference proceedings.}
\end{LARGE}

\vspace{2cm}
\textbf{Glenn Wouda}\\
e-mail: \verb|glenn.wouda@physics.uu.se| \\ 
Department of Physics and Astronomy,\\ Uppsala University,\\ 
Box 516, SE--751 20 Uppsala, Sweden    
\vspace{1cm}

\begin{center}
\textbf{Abstract}
\end{center}
I present a model with two scalar doublets and a softly broken $\ztwo$ symmetry, where only one of the doublets gets a vacuum expectation value and couples to fermions at tree-level. The softly broken $\ztwo$ symmetry leads to interesting phenomenology such as mixing between the two doublets and a charged scalar $\hp$ which can be light and dominantly decays into $W \gamma$. The model can also naturally reproduce an enhanced $\gamma \gamma$~signal of the newly observed Higgs boson at the LHC with mass 125 GeV.

\section{Introduction}
\label{intro}
The discovery of a new particle was announced by the ATLAS~\cite{ATLAShiggs} and CMS~\cite{CMShiggs} experiments at the Large Hadron Collider~(LHC) in 2012. Further analyses has shown that this new state indeed is a boson and its properties are very similar those of the predicted Higgs boson of the Standard Model~(SM)~\cite{ATLAS-CONF-2013-012}. This discovery is certainly one of the greatest successes of elementary particle physics, but there are many questions left to be answered and understood. One question is if it	 exists several Higgs bosons. Nature seems to have arbitrarily introduced several generations of fermions. There is no known reason for the existence of three generations of fermions, and perhaps a Higgs sector is composed out of several generations (multiplets) of Higgs fields as well. 

A simple and very versatile extension of the SM Higgs sector is the so called two-Higgs Doublet Model (2HDM). As its name suggests, the Higgs sector consists of two Higgs doublets. The potential reads
\begin{align}
\mathcal{V}   &= m_{11}^2|\Phi_1|^2  +m_{22}^2 |\Phi_2|^2
-[m_{12}^2\Phi_1^\dagger\Phi_2+{\rm h.c.}]
+ \nonumber\\
\frac{\lambda_1}{2}&|\Phi_1|^4  + \frac{\lambda_2}{2}|\Phi_2|^4 
+\lambda_3 |\Phi_1|^2 |\Phi_2|^2
+\lambda_4 | \Phi_1^\dagger\Phi_2 |^2
\nonumber\\
+\,\bigg\{ &\frac{\lambda_5}{2}(\Phi_1^\dagger\Phi_2)^2  
+\big[ \lambda_6 |\Phi_1|^2 
+\lambda_7 |\Phi_2|^2  \big]
\Phi_1^\dagger\Phi_2 +  {\rm h.c.} \bigg\} \, ,
\label{eq:Vpotential}
\end{align}
where $\Phi_{i}$ are $\grp{SU(2)}_L$-doublet, hypercharge $Y=1$, scalar fields. The index $i = 1,2$ is called the Higgs-flavour index. In general, the parameters $\lambda_{5,6,7}$ and $m_{12}^2$ are complex and will introduce a source of $\mathcal{CP}$ violation. The remaining parameters are real. The potential \eref{eq:Vpotential} has a global $\grp{U(2)}$ symmetry in Higgs-flavour space. One can define a basis in the space of $\grp{U(2)}$ transformations in terms of the vacuum expectation values (VEVs) of the $\Phi_{i}$'s 
\begin{equation}
 \vev{\Phi_1}^T = (0 , v_1) \,, \: \quad
 \vev{\Phi_2}^T = (0 , v_2 e^{\ii \xi}) \, .
\end{equation}
The VEVs fulfil $v_1^2 + v_2^2 = v^2 \approx 246$ (GeV)${}^2$ and a non-zero $\xi$ will spontaneously break $\mathcal{CP}$. We will from now on only consider $\mathcal{CP}$ invariant Higgs sectors. It is customary to define the parameter $\tb \equiv v_2/v_1$. However, due to the $\grp{U(2)}$ symmetry of \eref{eq:Vpotential}, $\tb$ is not a physical parameter until a specific Yukawa structure is imposed. We refer to \cite{Branco:2011iw} for a review of general 2HDMs.

In the absence of the parameters $m_{12}^2$ (soft terms) and $\lambda_{6,7}$ (hard terms) the potential \eref{eq:Vpotential} possesses a $\ztwo$ symmetry 
\begin{equation}
 \Phi_1 \to  \Phi_1  \,, \qquad \Phi_2 \to -\Phi_2 \,,
\end{equation}
which can be interpreted as a Higgs-flavour parity. It is important to not break this symmetry with hard terms in order to naturally avoid potentially large flavour changing neutral currents (FCNC) \cite{Glashow:1976nt,Paschos:1976ay}. This means that one assigns $\ztwo$ parities also to the fermions in such a way that only one Higgs doublet couples to every fermion species at most. For instance, we can make sure that only $\Phi_1$ couples to fermions. This is called a Type-I 2HDM. The Yukawa Lagrangian will break the $\grp{U(2)}$ symmetry of \eref{eq:Vpotential} and thus promote $\tb$ to a physical parameter at tree level.

The 2HDM particle spectrum consists of the $\mathcal{CP}$ even $h,H$, the $\mathcal{CP}$ odd $A$ and the charged $\hp$. The charged Higgs boson $\hp$ has been considered to be a ``smoking gun'' for physics beyond the SM. The $\hp$ will in most of the parameter space in models with a 2HDM structure decay dominantly into $\tau \nu$ if $m_\hp < m_t$ or into $tb $ if $m_\hp > m_t$. This is the case for the Minimal Supersymmetric SM (MSSM) and a large effort has been made to discover the $\hp$ in any of these channels, see e.g.~\cite{Abbiendi:2013hk}. Up to now, only~exclusion limits has been set on e.g. the branching ratio BR$(t \to H^+ b) $ see e.g.~\cite{Aad:2013hla}. As these limits are getting tighter, it is reasonable to go beyond these ``standard'' scenarios for Higgs physics beyond the~SM.

\section{The Stealth Doublet Model}
I will now present a model with a 2HDM structure, the Stealth Doublet Model (SDM). This model has been proposed and presented earlier in~\cite{Wouda:2010zz,Enberg:2013ara,upcoming} and has many new and unusual phenomenological features. By considering the 2HDM potential \eref{eq:Vpotential}, the following postulates are made: the $\ztwo$ symmetry is broken only softly and the physical basis is the one where only $\Phi_1$ has a VEV, $v_1 = v$, and couples to fermions. The Yukawa-Lagrangian at tree-level then reads in unitary gauge
\be
-\mathcal{L}_{\:\text{Yukawa}} = (m_f/v)\, \bar{\Psi}_f \Psi_f \: \Phi_{1} .
\ee
This is very similar to the starting point of the Inert Doublet Model (IDM) \cite{Barbieri:2006dq} with the difference that there the $\ztwo$ symmetry is kept exact at all scales. One can therefore interpret the SDM as the (continuous) generalization of the IDM. 

The soft breaking of the $\ztwo$ symmetry will prevent large FCNCs in a natural way since at very high energies, $E \gg |m_{12}|$, the $\ztwo$ symmetry is restored. A UV-completion of the SDM could be considered but here a bottom-up approach is adopted where only the effective theory at lower energies is studied. The formalism of soft symmetry breaking terms is a conventional way of parametrizing the ignorance of how the symmetry is broken and allows one to study low energy phenomena. This is standard practice in e.g.\ the MSSM. In order to have the $\ztwo$~symmetry softly broken, the potential parameters should fulfil the relations~\cite{Davidson:2005cw}
\begin{align}
(\lambda_1 - \lambda_2)\left[ \lambda_{345} \right. &  \left. (\lambda_6+\lambda_7)-\lambda_2\lambda_6-\lambda_1\lambda_7\right] \nonumber\\
&-2(\lambda_6-\lambda_7)(\lambda_6+\lambda_7)^2 = 0, \nonumber\\
 (\lambda_1 - \lambda_2)m_{12}^2+&(\lambda_6+\lambda_7)(m_{11}^2-m_{22}^2) \neq 0,
\end{align}
which define the parameter space of the model. One way to fulfil these relations is to choose $\lambda_2 = \lambda_1$ and $\lambda_7 = \lambda_6$. For the minimization conditions of the potential and mass relations of the scalars in SDM, the reader is referred to \cite{Wouda:2010zz,Enberg:2013ara,upcoming}.

Since it is only $\Phi_1$ that obtains a VEV and couples to fermions, the states $A$ and $\hp$ are composed out of $\Phi_2$ components solely and are fermiophobic. They are therefore immune to practically all indirect constraints from low-energy flavour observables. The mass eigenstates $h$ and $H$ ($m_H > m_h$) are mixtures of the real component interaction eigenstates from $\Phi_1$ and $\Phi_2$
\be
 H =  \phi_1\cosa + \phi_2 \sina, \:\:\:\: h = - \phi_1 \sina + \phi_2 \cosa .
\ee
The mixing is parametrized by the angle $\alpha$ which is introduced in order to diagonalize the mass matrix in the $\mathcal{CP}$ even sector. The input parameters are taken to be $m_h, \ m_H, \ m_A, \ m_{H^\pm},  \ \sina , \ \lambda_3$ and $ \ \lambda_7 $. All the other potential parameters can be solved for using the minimization conditions, inversion of the mass relations and the conditions for soft $\ztwo$ breaking. The parameter $\sina$ is proportional to $m_{12}^2$ and can be interpreted as the amount with which the $\ztwo$ symmetry is broken. The parameters $\lambda_3$ and $\lambda_7 $ enter the $hH^+H^-$ and $HH^+H^-$ couplings and therefore influence the widths $\Gamma_{h/H \to \gamma \gamma} $.

\begin{figure}[t]
\centering
\includegraphics[scale=0.35]{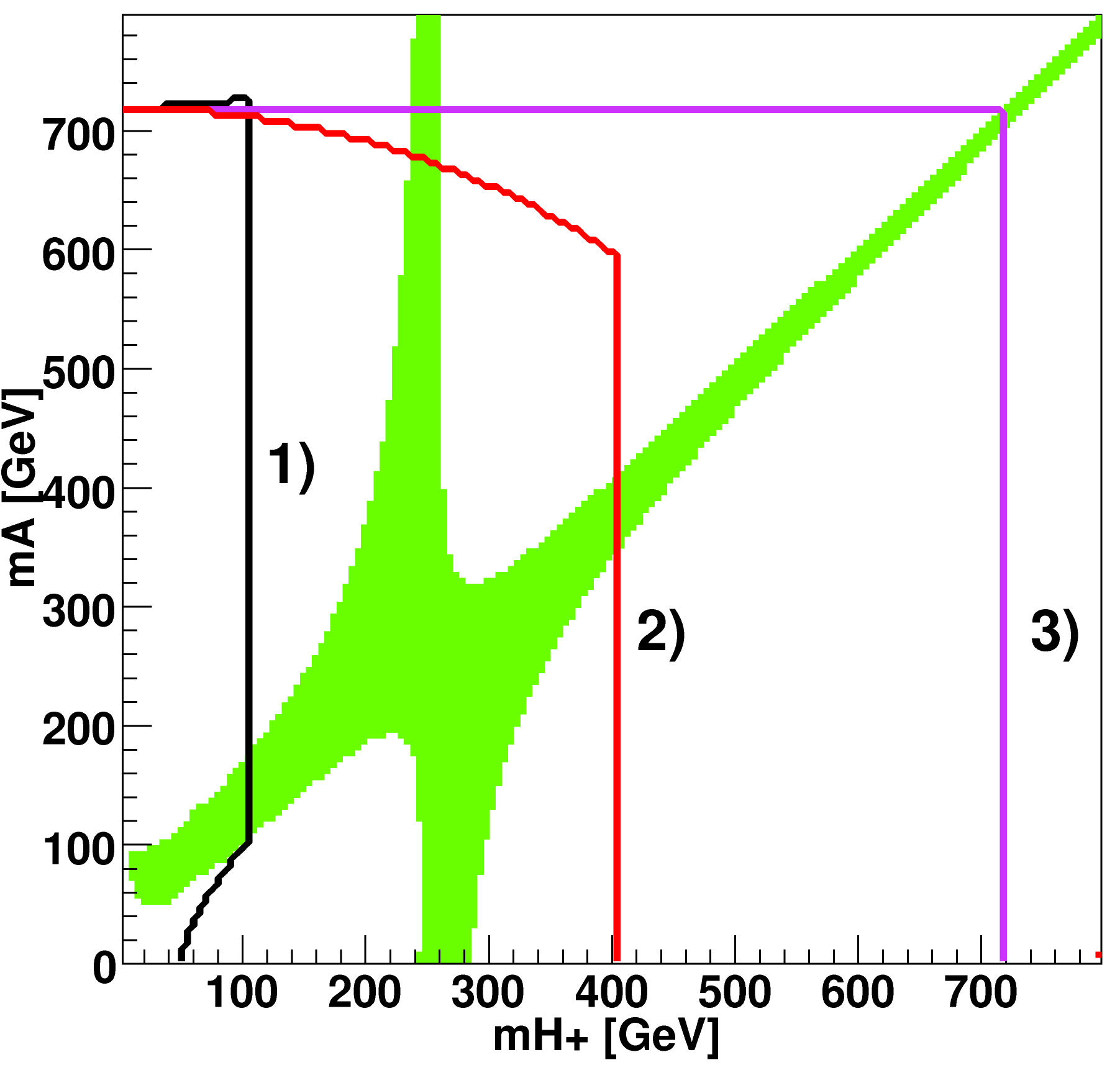}
\caption{The green area shows the $S$ and $T$ values which falls inside the 90\% C.L. ellipse of Figure 10.7 in \cite{PhysRevD.86.010001}. To the left of the bands are the allowed regions by the theoretical constraints for different values of $\lambda_3$: 1) $\lambda_3 = 0$, 2) $\lambda_3 = 4(m_\hp/v)^2 $ and 3)~$\lambda_3 = 2(m_\hp/v)^2 $. Here $m_h = 125 $ GeV, $m_H = 300$~GeV, $\sina = 0.866$, $\lambda_2 = \lambda_1$ and $\lambda_7 = \lambda_6$ is used.}
\label{fig-1}       
\end{figure}

Constraints on the model are considered from Electroweak precision tests (EWPT) -- the oblique $S$ and $T$ parameters. Constraints from theory: a potential bounded from below, tree-level unitarity and perturbativity of the four-scalar vertices, are also considered. These constraints were calculated using the software \THDMC~\cite{Eriksson:2009ws} and in \reffig{fig-1} examples of allowed regions in parameter space are shown. One should note that the $S$ and $T$ parameters have no explicit dependence on the $\lambda_i$ parameters and that $m_A \gtrsim m_\hp + 50$~GeV must be fulfilled for light $m_\hp$.
\begin{figure}[bth]
\centering
\includegraphics[scale=0.35]{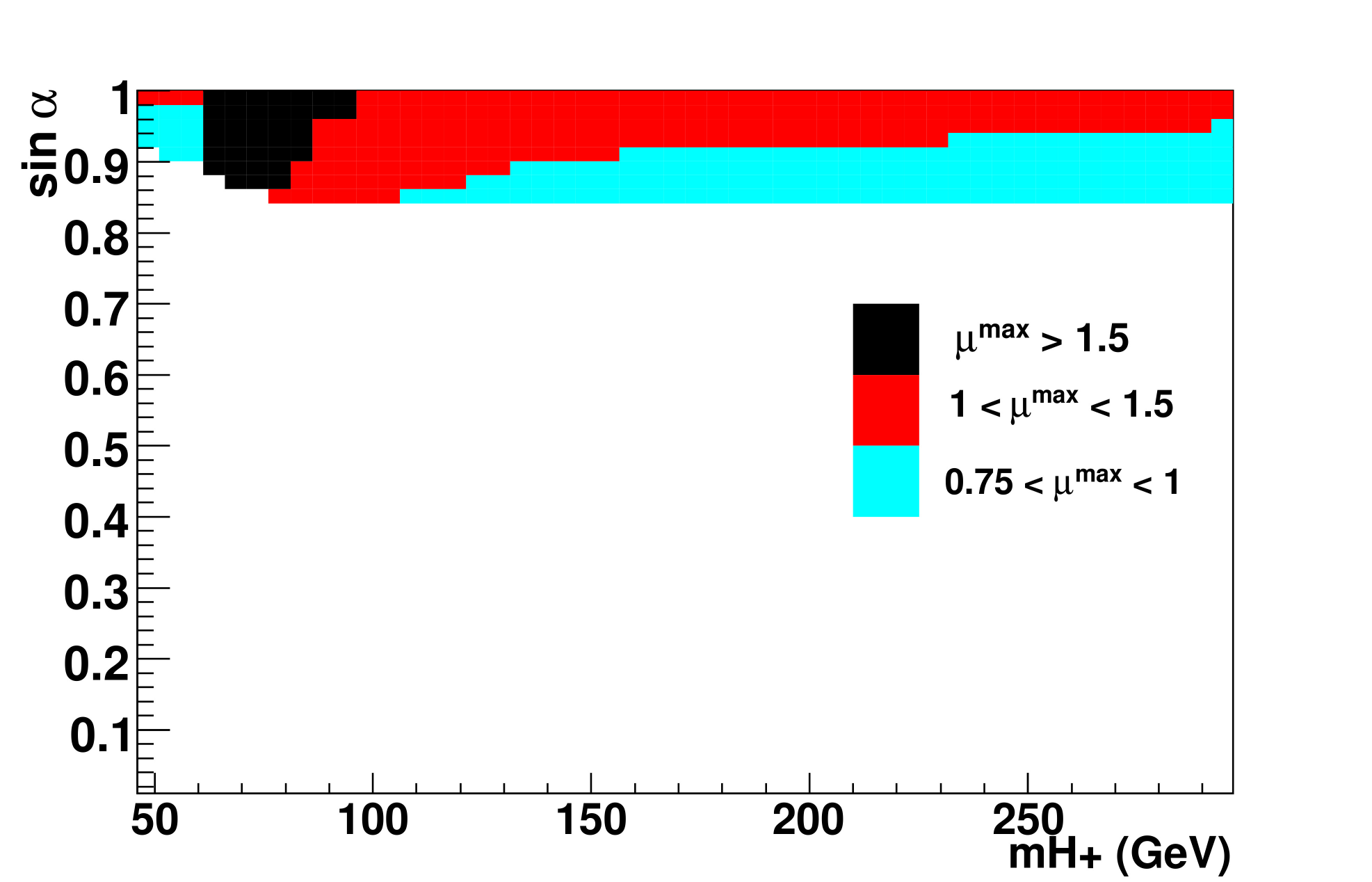}
\caption{The maximally found signal strength $\mu^{\text{max}}$ for $h\to\gamma \gamma$. Here, $m_h = 125$ GeV, $m_H = 300$ GeV, $m_A = m_\hp + 50$~GeV.}
\label{fig-4}       
\end{figure}

\section{The SDM and the new observed state}
This model has two candidates for the discovered Higgs~boson $\hcal$, $ m_\hcal \approx 125$ GeV, the $\mathcal{CP}$ even states $h$~and~$H$. Either $h$ is the observed $\hcal$ and the heavier $H$ remains to be discovered. Or the heavier $H$ is the observed state and the lighter $h$ has not been detected yet.

By scanning over the model parameter space, regions which are compatible with EWPT and theoretical constraints and earlier searches for Higgs bosons are found. An additional requirement is that compatible signal strengths for the $\gamma \gamma$ and $ZZ$ channels are fulfilled \cite{ATLAS-CONF-2013-012,ATLAS-CONF-2013-013}. The results of the scans are that if $h=\hcal$ then $\sina \gtrsim 0.8$ and if $H=\hcal$ then $\sina \lesssim 0.5$. \reffig{fig-4} shows the maximally found signal strength $h\to\gamma \gamma$, see \cite{Enberg:2013ara,upcoming} for details.

\section{Decays of the $A$ and $H^\pm$ bosons}
\begin{figure}[tbh]
\centering
\begin{tabular}{ccc}
\includegraphics[scale=0.40]{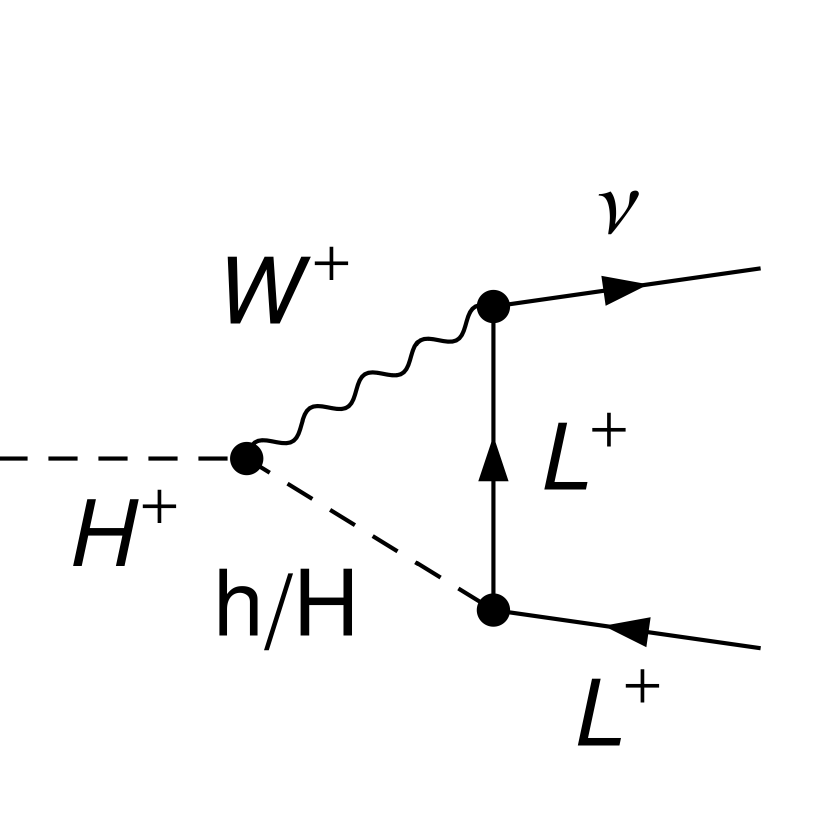}& \includegraphics[scale=0.40]{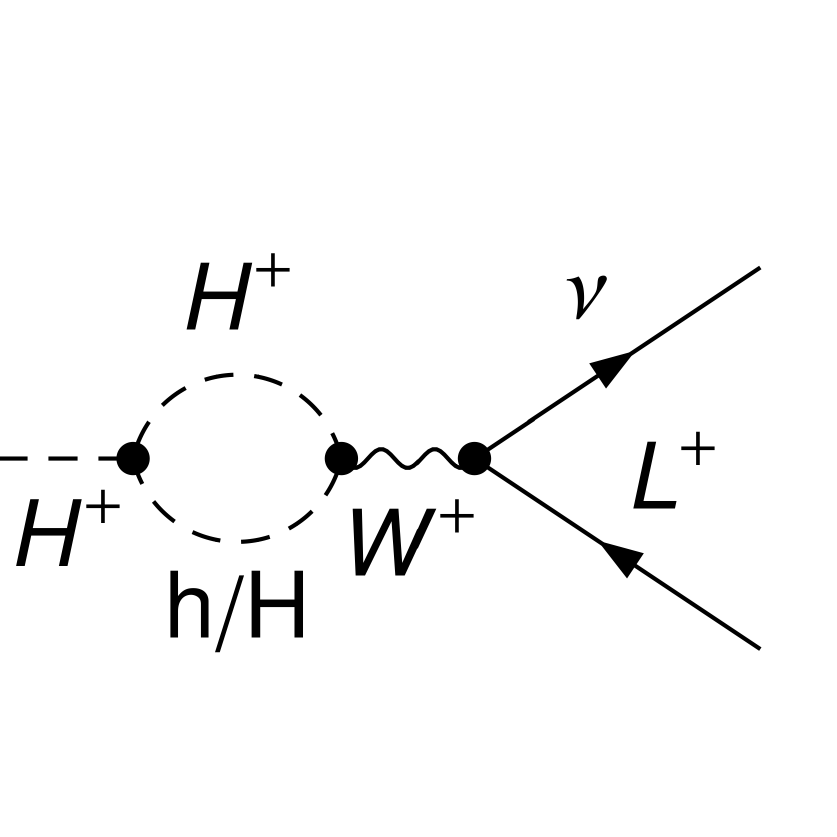} &\includegraphics[scale=0.40]{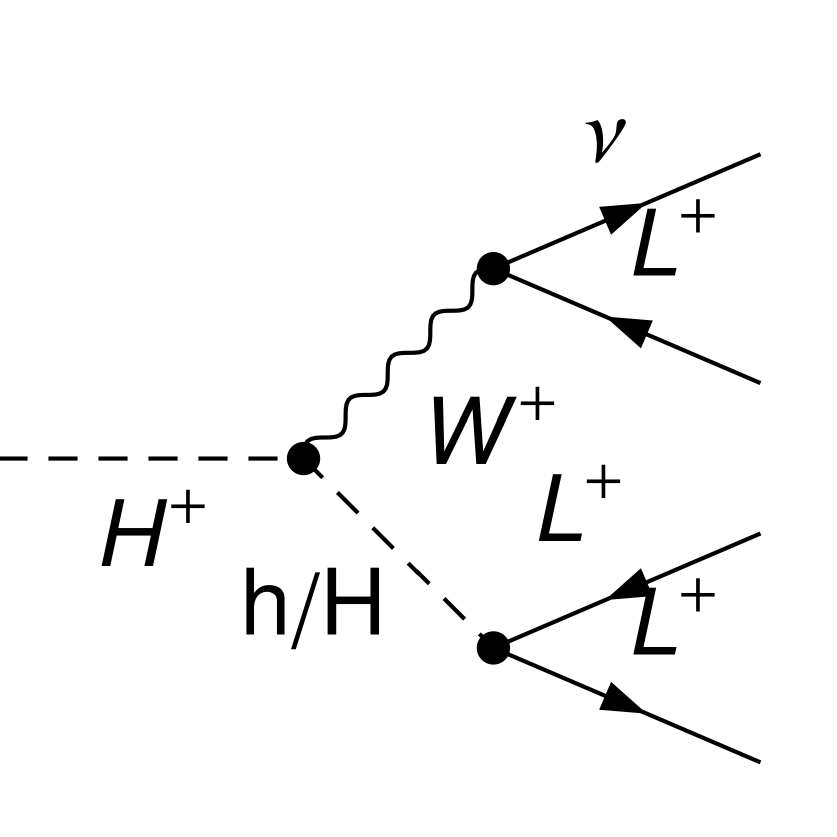}
\end{tabular}
\caption{Examples of Feynman diagrams for $\hp \to 2$ or $4$ fermions.}
\label{fig-2}       
\end{figure}

Due to the mixing, it is possible for the $A$ and $\hp$ bosons to couple to fermions. In particular, they can couple to two fermions at one-loop level. Another possibility for them to have fermionic decay modes is via (off-shell) $h,H$ and $Z$ (for $A$) or $W$ (for $\hp$) see \reffig{fig-2} for sample Feynman~diagrams. For the charged scalar the decays $\hp \to W \gamma$ and $\hp \to W Z$ are also possible at one-loop level. The tree-level processes were calculated using \MadG { }\cite{Alwall:2011uj} and the loop processes with \FC { }\cite{FormCalc} and associated packages. The effect of the $W$-width in $\hp \to W \gamma$ was calculated using the Smeared mass unstable particle model \cite{Kuksa:2009de}. Taken this into account, it was shown that $\hp \to W^* \gamma$ dominates over $\hp \to f\bar{f}^\prime$ for $m_\hp < m_W$. 

\begin{figure}[t]
\centering
\includegraphics[scale=0.85]{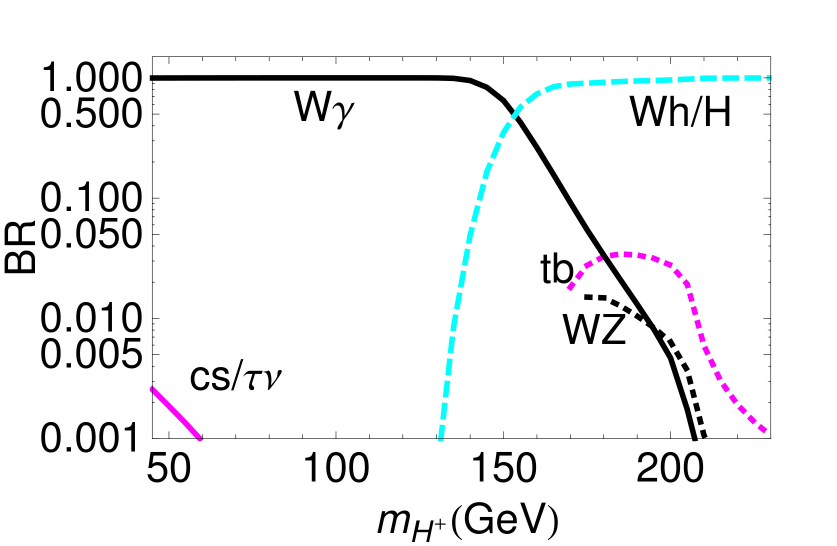}
\caption{Branching ratios for the $\hp$ boson. In this figure $m_h$ = 125 GeV, $m_H$ = 300 GeV, $m_A = m_\hp + 50$ GeV, $\sina = 0.866$, $\lambda_3 = 2 (m_\hp /v)^2$, $\lambda_2 = \lambda_1$ and $\lambda_7 = \lambda_6$.}
\label{fig-3}       
\end{figure}

Below the $\hp \to W^* h$ threshold, the mode $\hp \to W \gamma$ always dominates if $\hp$ is the lightest scalar, see \reffig{fig-3} for an example. This is the first model where $W \gamma$ could be the naturally dominating decay mode for charged scalars despite being loop suppressed. The branching ratios for the $\hp \to WZ$ and the standard $\hp \to tb$ modes are typically of the order a few percent, see \reffig{fig-3}.

The decay modes of the $A$ boson are standard, if it is the lightest scalar $b \bar{b}$ dominates and $Z^* h$ or $W^* \hp$ otherwise.

\section{Outlook}
Since $A$ and $\hp$ are fermiophobic, their standard production mechanisms in other models are not applicable here. For instance, the $\hp$ will not be produced in $t$~quark decays even if light. Instead Drell-Yan type processes such as $q\bar{q} \to Z^*/\gamma^* \to H^+ H^-$, $q\bar{q}^\prime \to W^* \to H^+ A$, or via (possible off-shell) $h$ and $H$ from gauge boson fusion are to be considered. How to detect light charged scalars in the $W\gamma$ channel remains to be investigated.

\subsection*{Acknowledgements} I would like to thank the organizers for the opportunity to present this work with a poster at the conference and in this article. Collaboration with Rikard Enberg (Uppsala University) and Johan Rathsman (Lund University) is acknowledged. 

\bibliographystyle{woc}
\bibliography{lopsided}

\end{document}